\documentclass[prl,final,showpacs,superscriptaddress]{revtex4}
\usepackage{bm}
\usepackage{times}
\usepackage{amsmath}
\usepackage{amsfonts,amssymb}
\usepackage{graphicx}

\begin{document}
\large
\begin{titlepage}
\title{Photon Pickup by Intense Poynting Flows}
\author{David Eichler}
\affiliation{Physics Department, Ben-Gurion University,
Beer-Sheva}

\begin{abstract}
   It is suggested that a Poynting flux-dominated outflow with a
sufficiently strong magnetic field can pick up hard X-ray photons
when the magnetic field is of sufficient strength. The zeroth
generation  X-rays pair produce, and the pairs radiate extremely
energetic first generation photons that could be detected by
extensive air shower arrays and/or MILAGRO if they escape the
production site intact. Giant flares from magnetars may thus yield
bursts of UHE photons. GRB-associated Poynting flows may be
unstable to pair production near their source, and their energy
rapidly converted to pairs.

\end{abstract}
\maketitle
\end{titlepage}

\section{Introduction}

There is growing evidence that soft gamma ray repeaters (SGR's)
and anomalous X-ray pulsars are  objects with ultrastrong magnetic
fields, $10^{14.5}$ G or more \cite{DT,TD,{WO}}. Giant flares from
SGR's appear to be caused by a reconnection or reconfiguration of
an  ultrastrong magnetic field, in a manner similar to the
mechanism that causes solar flares. These flares have two stages:
a) the bright phase, lasting $\sim 10^{-1}$s, when the energy is
presumably released, b) a period lasting several minutes during
which time the remaining plasma is apparently trapped on closed
field lines.  More than $10^{44}$ ergs may be released in giant
flares, while smaller, more frequent bursts field between
$10^{38}$ and $10^{42}$ ergs.

A rapidly rotating neutron star with a magnetic field of at least
$3 \times 10^{14}$ G  was proposed by Usov \cite{US} as being the
source of gamma ray bursts (GRB's). One indication (of sorts) is
that synchrotron models of afterglow frequently invoke a rather
large fraction of the total fireball energy in Poynting flux to
obtain good fits (e.g. \cite{LW}). Other models \cite{T94,LPB}
explicitly invoke Poynting flux as being the original form in
which GRB energy is expelled by the compact object. The
 rate at which  magnetars form can be estimated by dividing the number
in the Galaxy ($\sim10$) by their age ($\sim 10^4$ yr), and the
rate so estimated,  once per $\sim 10^3$ years per galaxy, is
consistent with what is needed to produce GRB's. Perhaps magnetars
are the fossil remnants of GRB's.

It would be useful to identify a qualitative consequence of
ultrastrong magnetic fields, one that would not exist to any
degree were the fields much smaller.  In this letter, it is
suggested that pair production by photons in a magnetic field
could be just such a consequence and could result in the emission
of ultrahigh energy (UHE) photons.   The basic point is that such
coupling is possible in a magnetic field only when the product of
the photon energy $\epsilon_{\gamma}$ and field strength
$B_{\bot}$ normal to the photon path  satisfies

\begin{equation}
\epsilon_{\gamma}B_{\bot} \ge 2 \times 10^{18}eV-G.
\label{criterion}
\end{equation}
 Thus, when the field
strength exceeds $10^{14}$G, a photon above 20 KeV (routine for
typical astrophysical thermal X-ray emitters)  suffices.
(Hereafter the subscript in $B_{\bot}$ will be dropped.)

We will assume  in this letter  without rigorous proof that, in a
flare-like event,  plasma can be ejected from a magnetar or
similar object to infinity, much like plasma is ejected from the
Sun by solar flares, but with a very large Lorentz factor
($\Gamma\ge 10^4$) characteristic of the large Alfven Lorentz
factor associated with tenuous plasma in a strong magnetic field.
At this $\Gamma$, the plasma moves essentially at the speed of
light, to within one part in $10^8$.  There is little relative
motion between the photon and the plasma over a hydrodynamic
timescale, which, in the frame of the fluid, is much shorter than
the crossing time  measured in the lab frame.

\section{Basic Considerations}
Consider a strongly magnetized compact object (SMCO),  e.g.
neutron star, collapsar or accreting black hole of radius R,  with
a field strength B, $B\gg B_{QED} \equiv 4.4 \times 10^{13}$G.
Because the gravitational field allows a negligible thermal scale
height, the plasma density well above the surface is determined
purely by electrodynamics, and is of order $B/4\pi R\beta$, where
$\beta c$ is the velocity of the charge carriers. The
characteristic current density associated with the field is of
order $c B/4\pi R$, and the minimum proton or electron number
density associated with this charge density, $n=j/e\beta c$, is $n
\sim B/4\pi R e\beta c \sim 2 \times 10^{17}B_{15}R_6/\beta\, \rm
cm ^{-3} $ where numerical subscripts of any quantity refer to
powers of ten by which the quantity is to be multiplied when
expressed in cgs units. If the current is carried by protons, then
the energy per current-carrying proton, $B^2/8\pi n$, is of the
order of
\begin{equation}
\sigma m_pc^2 \equiv eBR =
 10^{14.5}B_{15}R_6 m_pc^2.
\end{equation}
For a magnetar, the energy per current-carrying proton can be as
high as $10^{14.5}m_pc^2$. If an overweight,  "failed" or
spun-down magnetar collapses to a black hole \cite{VS}, then, just
prior to completing this collapse, it would have a field of about
an order of magnitude higher, and  $ \sigma \sim 10^{15}$. (In
practice, a single species plasma would result in extremely high
electric fields and the plasma is likely to be quasi-neutral. In
this case, the multiplicity $\xi$, i.e. the ratio of the actual
plasma mass density to the minimum needed to provide the curl of
{\bf B}, is likely to be  greater than unity. In pulsars, the
charge multiplicity is unknown, but could be as high as $10^4$  or
more \cite{HA}, and  if, as will be assumed here, the current
carriers include ions,  the mass multiplicity is even less
certain.) If a significant fraction of this magnetic energy
density  is stored in a solenoidal component of this field, it
could be released by reconnection and current dissipation.

To simplify geometric considerations, consider a uniform magnetic
field {\bf B}=$10^{15}B_{15}${\bf b}G, where {\bf b} is the unit
vector in the field direction,  and a frame that moves
perpendicular to {\bf B} with Lorentz factor $\Gamma$ relative to
the frame of the compact object, hereafter called the lab frame.
In the zero electric field frame  (ZEFF), the magnetic field
strength is

\begin{equation}
B'= B/\Gamma
 \label{B'}.
\end{equation}
This basic fact is used repeatedly below. The Lorentz factor of
the particle in the ZEFF will be denoted by $\gamma'$. Note that
 criterion  (\ref{criterion}) is not only Lorentz invariant, as
could be expected intuitively, but also, in light of equation
(\ref{B'}), independent of $\Gamma$ to within geometric factors.
Ultrarelativistic motion of the plasma, while lowering $B'$
relative to B, raises $\epsilon'^{(o)}$ relative to
$\epsilon^{(o)}$, the lab frame energy of the incident photon,  by
about the same factor.

Another necessary condition for pair production is that the photon
energy in the ZEFF exceed $2m_ec^2$.  Photons this energetic are
not typically emitted by thermal processes in astrophysical
situations.  However, if the ZEFF is sufficiently
ultrarelativistic in the frame of the photon source, this is not a
problem for X-rays, for then the photon as seen in the frame of
the plasma is Doppler-boosted well above the pair production
threshold.

\section{Photon Pick-up}
  A photon with lab frame energy $E_{ph}$ has an energy in the
frame of the piston of $E'_{ph}=E_{ph}\Gamma $  to within a
geometric factor. Photons satisfying $E_{ph}'B'\ge 2\times
10^{18}$eV-Gauss are converted to pairs. The pairs   emit
synchrotron radiation at a frequency of $2\times 10^7
\gamma'^2B_0'$ in the local comoving frame where $B_0'$ is the
ZEFF field strength in Gauss. As they slow down, they emit an
increasing number of photons per decrease in ln $\gamma'$.
Assuming that the pair has  a pitch angle of 90 degrees, then, in
the  classical approximation,
\begin{equation}
d\gamma'/dt'= -2\gamma'^2\sigma_T B'^2/8\pi m_ec \label{gamma'}
\end{equation}
and
\begin{equation}
d\theta'/dt' = eB'/\gamma'm_ec.
 \label{theta}
 \end{equation}
Choosing  $\theta = 0$  to be  the starting direction of the pair,
i.e. opposite to the direction of the plasma, and assuming the
initial $\gamma'$ to be very large,  we
 can write
 the solution to equations (\ref{gamma'}) and (\ref{theta}) as
 \begin{equation}
\gamma'= \left(\frac{3c}{4r_o\omega'_c\theta'}\right)^{1/2}.
 \end{equation}
The characteristic energy of the emitted photons $\epsilon'$ is
given by
\begin{equation}
\epsilon' =\frac{3c\hbar}{4 r_o \theta'}
 \label{ep'}.
 \end{equation}
The lab frame energy $\epsilon$ is given by
\begin{equation}
\epsilon = \Gamma (1 - \beta cos\theta')\epsilon'.
 \label{epsilon}
 \end{equation}
The spectrum of emitted photons as seen in the lab frame is easily
calculated by noting that the rate of photon emission per unit
proper time in the plasma frame is constant:
\begin{equation}
dN/dt' = R
 \end{equation}
where N is the cumulative number of photons that have been emitted
and  $R \sim \sigma_T c (B'^2/8\pi \hbar \omega'_c)$, and
\begin{equation}
N= A \theta'^{1/2}\label{N}
 \end{equation}
 where $A=
 \frac{2R}{\omega_c}\left(\frac{3c}{4r_o\omega_c}\right)^{1/2}$.

Writing the total energy spectrum in the lab frame is complicated
by the fact that (\ref{ep'}) and (\ref{N}) combine to form a
transcendental, non-monotonic relation between N and $\epsilon$.
However, it is easy to write $\epsilon$ as a function of N by
using  (\ref{N}) and (\ref{ep'} in (\ref{epsilon}):
\begin{equation}
\epsilon = \frac{3\hbar c}{4 r_o}\Gamma[1-\beta
cos([N/A]^2)]/[N/A]^2.
\end{equation}
The solution can then be "turned on its side" to obtain
$dN/d\epsilon\equiv \sum(dN/d\theta')/(d\epsilon/d\theta')$ by
summing the contributions of the individual segments for which N
is monotonic in $\epsilon$. From here on, $\beta$ will be taken to
be essentially unity. In figure 1, the individual components of
the spectrum are plotted for the first 10 gyrations, gyration by
gyration, in the delta function approximation [see eq. (7)]. The
total spectrum at most energies (not shown) is dominated by the
individual peaks and appears quite similar to the graph in the
figure.

\begin{figure*}
\includegraphics{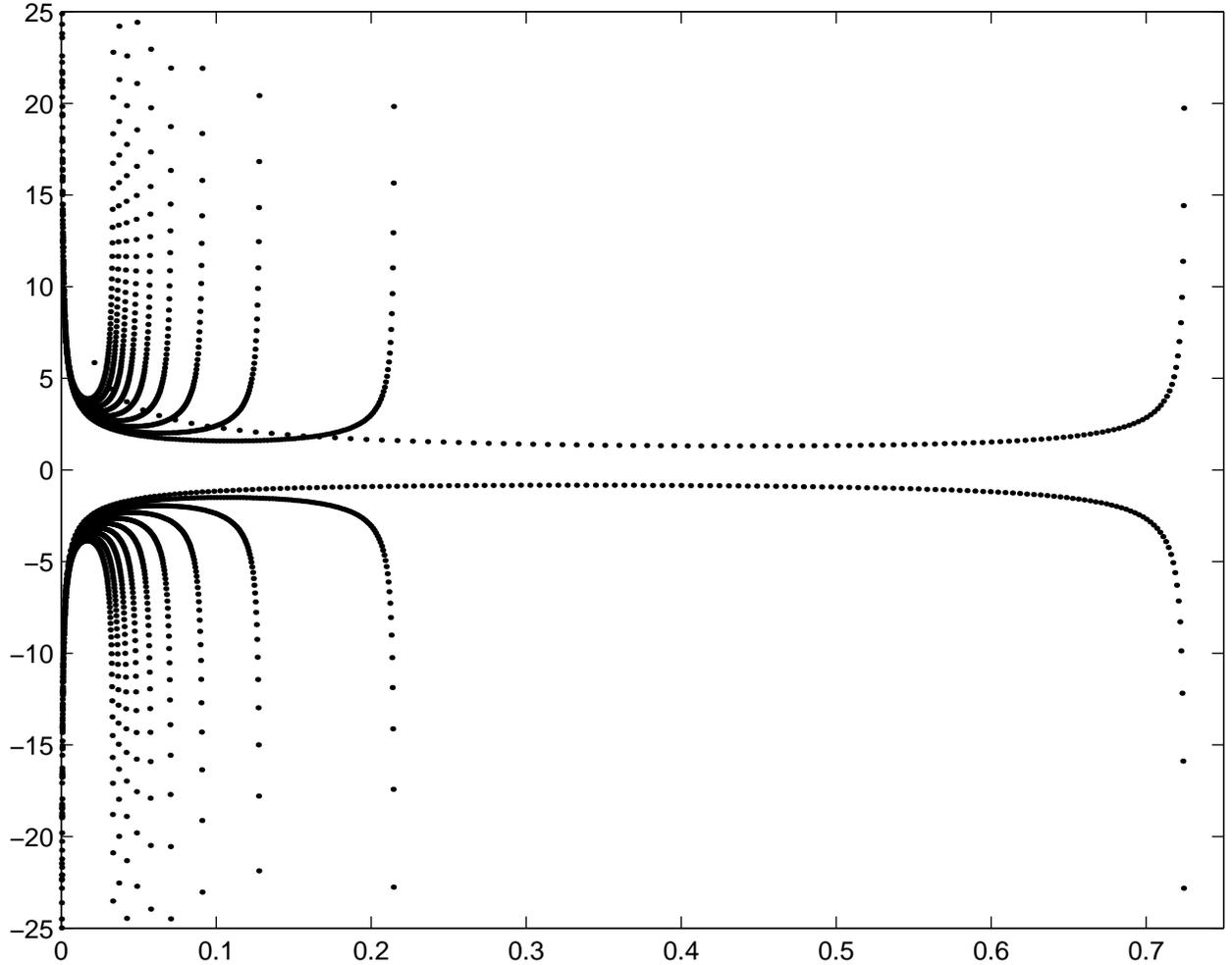}
\caption{The differential  spectrum dN/d$\epsilon$ of gamma rays
emitted by a pair that is produced by a picked-up photon. For
simplicity, it has been assumed that the instantaneous spectrum is
a delta function at energy $\epsilon'$ given by
equation((\ref{ep'}); the actual peaks would be somewhat broader.
The pair begins moving in the backward direction ($\theta = 0$)
relative to the motion of the fluid element. The lab frame energy
of the emitted photons rises as the pair gyrates toward the
forward direction, and, at $\theta ^*= 2\pi n + 2.33 \sim 2\pi n +
3\pi/4 $, begins to decline. To graphically separate the rises and
declines, the spectral contributions from the declining phases are
plotted as negative numbers, i.e. -dN/d$\epsilon$. The unit of
energy on the x axis is $3\Gamma m_ec^2/4\alpha = 52.5 \Gamma
$MeV. The interval between adjacent dots denotes $1.2^o$ of
gyrophase, 300 per gyration. The y axis is left arbitrary but the
total number of emitted photons is easily estimated as $\gamma'
m_ec^2/\epsilon'$. The divergence of dN/d$\epsilon$ at $\theta ^*$
is due to the vanishing of $d\epsilon/dt$, but the total area
under the peak is finite.} \label{fig1}
\end{figure*}


 At low energy, the spectrum is essentially
 \begin{equation}
d N/d\epsilon \propto \epsilon ^{-3/2}
 \end{equation}
 which is a familiar result for the time-averaged emission of
 decelerating, energetic particles. At high energy, most of the
 energy in the first generation photons is emitted in a very narrow peak near

\begin{equation}
\epsilon^{(1)} = [1- cos(3\pi/4)]\Gamma m_ec^2/\pi \alpha .
\label{20}
\end{equation}
 The lab frame energy is thus
\begin{equation}
 \epsilon^{(1)}\sim 38 \Gamma MeV
 \label{21}
\end{equation}
which, for the magnetosonic value of $\Gamma$,  $\Gamma =
(\sigma/\xi)^{1/2}$, can exceed $10^{15}(\sigma_{15}/\xi)^{1/2}$
eV.  Note that the lab frame energy depends only on the Lorentz
factor of the fluid element in which it is produced.

Photons this energetic survive in the high $\Gamma$ outflow, which
has a much lower field than the lab frame,
provided that \begin{equation}
 \epsilon'B' \sim 38 MeV B/\Gamma \le
2\times 10^{18}eV-G .
 \label{ppthr}
\end{equation}
  This is satisfied when $\Gamma \ge (B/5
\times 10^{10}$Gauss). For magnetar strength fields ($B \sim 5
\times 10^{14}$G), $\Gamma$ must exceed $ 10^{4}$, and the
surviving photon energies must therefore exceed 0.5 TeV.
 The velocity difference between the fluid element
and the photon is exceedingly small, and they move together.  If
the fluid element is ejected to infinity from the magnetosphere,
in analogy to fast streams from solar flares, then the photon
moves with the host fluid element out to infinity, where the field
is too weak for further pair production.

If, on the other hand, the photon is produced within  a fluid
element  where equation (\ref{ppthr}) is not satisfied, the pair
cascade continues. The second generation pairs would have ZEFF
energy $\epsilon' \sim 3m_ec^2/2\pi \alpha $ and the second
generation photons they emit would have a lab frame energy of
\begin{equation}
\epsilon^{(2)} = \left({\frac{3}{2\pi \alpha}}\right)^2
\left(\frac{\Gamma B'}{B_{QED}}\right)m_ec^2.
\end{equation}
  In general, this is not enough to escape the
thermal photon field of the compact object (see below), and a pair
cascade should degrade the   energy of the higher generation
photons down to tens of MeV.



\section{Observational signatures}

   Let us now consider this mechanism in the context of flares from
SGR's. Our approach is to recognize the complexity of magnetic
field annihilation and merely assume without detailed argument
that magnetospheric motion is generated with a Lorentz factor of
about the Alfven Lorentz factor $[B^2/(8\pi\rho_o)]^{1/2}\sim
(\frac{\sigma}{\xi})^{1/2}$.  The value of $\sigma$ in a magnetar
magnetosphere can exceed   $10^{14}$, hence the Alfven Lorentz
factor, ${\sigma}^{1/2}$,  can exceed $\xi^{-1/2}10^7$. By
equation (\ref{21}), each of these   first generation photons can
result in $2N_{\gamma}$
 photons emitted at near $38 \sigma'^{1/2}\xi ^{-1/2}$MeV, where $N_{\gamma}$ is given by
\begin{equation}
N_{\gamma} \sim \left[\gamma'(\theta')
m_ec^2/\epsilon'(\theta')\right]_{\theta'=3\pi/4} \sim
\alpha^{1/2}(B'/B_{QED})^{-1/2}.
\end{equation}

A typical SGR has a persistent soft X-ray luminosity of order
$10^{35}$ erg/s, implying a quiet time magnetospheric photon
density of order $10^{21}$ photons/cm$^3.$ If each zeroth
generation photon results in $N_{\gamma}$ photons of $10^{15}$ eV,
up to $10^{24}N_{\gamma}$ ergs/cm$^3$ result in first generation
photons. For reasonable parameters, this can dominate the flare
energy ($10^{25}$erg/cm$^3$ is a typical value for the latter in
the case of large flares), implying that this mechanism may
actually absorb most of the flare energy. We suggest that this
allows  for the possibility of flares that are quiet in soft gamma
rays, while most of the energy is in the form of much more
energetic photons.

The UHE photons must still contend with thermal X-ray photons from
the magnetar surface, most of which pass easily through the
magnetic field. Assuming, somewhat generously, that the
X-radiation  is  half a black body photon field extending $5
\times 10^5 $cm, with temperature kT=0.5 KeV, then the $\gamma -
\gamma$ opacity felt by UHE gamma rays, which goes as
\begin{equation}
\kappa_{\gamma,\gamma}= (\pi^2/3)
\alpha^3r_o^{-1}\left(\frac{kT}{m_ec^2}\right)^3\left(\frac{(m_ec^2)^2}{\epsilon
kT}\right) ln\left(0.117\frac{(m_ec^2)^2}{\epsilon kT}\right)
\end{equation}
\cite{GS},  requires the photons to be above several TeV in energy
in order to escape the thermal photons from the magnetar surface.

The photon energy for first generation photons of
$\epsilon^{(1)}\sim 10^{15}$ eV is large enough to be detected as
an extensive airshower, and could be detected with surface arrays.
The naturally wide-angle fields of view of such detectors make
them especially suitable for monitoring the sky for occasional
bursts. Because outward expulsion of  plasma is necessary to
escort the photons through the otherwise high magnetic fields, we
predict that UHE photons would emerge only during the initial hard
spike of a giant flare, which lasts at most  tenths of seconds,
and is usually associated with an escaping pair plasma.

    Similarly, MILAGRO could pick up bursts of photons above 1
TeV. It is worth emphasizing that bursts from magnetars are rare,
the largest bursts occur only once per 20 years or so, and
positive detections at these energies might therefore also be
rare. In contrast to  most GRB's, which are  undetectable by
MILAGRO due to attenuation by pair creation over cosmological
distances, the occasional giant SGR flare, which is Galactic,
could very well be detectable.

Finally, we consider the possibility of detecting similar bursts
from nearby galaxies with MILAGRO. Making the best-case assumption
(among the wide range allowed by the uncertainty in $\Gamma$) that
the UHE photons emerge at $\sim 3 $TeV,  a burst of $\sim 10^{44}$
ergs would yield $\sim 10^{43.5}$  photons. If the detecting area
is   $10^4 A_{4}$ $m^2$, and assuming two or three simultaneous
photons constitute a detection, then photons are detectable out to
$\sim 3 A_{4}^{1/2}$ Mpc. Thus it might be worthwhile to monitor
for  directional coincidences with nearby galaxies within this
range.

 {\it GRB's:} If GRB's originate from Poynting flows out of strongly
magnetized compact objects (SMCO's), presumably rapidly rotating
neutron stars or black holes, then, close to the SMCO, the field
exceeds $B_{QED}$.  We now argue that such a flow, if surrounded
by hot plasma from an accretion, disk, corona, wind, or collapsing
core of the host star, is subject to the above mechanism of photon
pick-up and is unstable to the formation of a fireball. Zeroth
generation gamma rays of order $\epsilon^{(0)} \sim m_e c^2$ would
be emitted by the surrounding plasma and any gamma ray produced
within the Poynting outflow would have a good chance of scattering
off the wall and back into the outflow, while retaining an energy
of order $m_ec^2$. It would pair produce and emit more high energy
gamma rays, which would also pair produce. As long as
$B/B_{QED}\ge \Gamma \gg 1$, the electromagnetic energy can be
converted to pair energy in several hydrodynamic crossing times.
The scales at which this process takes place are less than
$10^8$cm, and the compactness parameter is sufficiently high that
most of the pairs would eventually annihilate.  The final output
would then be prompt gamma rays peaking at energies just below the
pair production threshold, quite like the thermal component of
GRB's. It would be hard to distinguish different scenarios (e.g.
simple two-photon pair production) for converting Poynting flux to
pair energy at the most compact scales of GRB's. In fact, other
pair production mechanisms (e.g. $\gamma-\gamma$) could continue
out to larger scales. We argue only that the strong magnetic field
presents at least one scenario that triggers the conversion of
Poynting flux to pair energy very early in the life of the GRB
eruption.

To summarize, we have suggested a  process for energetic photon
emission that  can happen only if $B \gg B_{QED}$. We suggest the
possibility of extremely energetic photons from giant magnetar
outbursts in the Galaxy, and perhaps nearby galaxies,  which would
be detectable with extensive air shower arrays.  However, because
significant disruption of the magnetosphere is necessary for such
energetic photons to escape, the events should be rare within a
single galaxy and their detectability  may rely heavily on the
persistent air time and large fields of view of extensive air
shower array detectors. A characteristic signature of the
mechanism would be a very hard spectrum with an abrupt high energy
cutoff.

We note that extreme Poynting flux from GRB's can be converted to
a pair fireball very close to the source by the same mechanism.

 I gratefully acknowledge very  helpful discussions with  Y. Lyubarsky,
and C. Thompson. This research was supported by an Adler
Fellowship administered by the Israel Science Foundation, by the
Arnow Chair of Astrophysics, and by the Israel-U.S. Binational
Science Foundation.

\end{document}